\documentclass[graybox]{svmult}
\usepackage{mathptmx}       
\usepackage{helvet}         
\usepackage{courier}        
\usepackage{type1cm}        
\usepackage{makeidx}         
\usepackage{graphicx}        
\usepackage{multicol}        
\usepackage[bottom]{footmisc} 
\usepackage[authoryear,sort&compress]{natbib} 

\makeindex             

\begin{document}

\title*{Extreme Star Formation}
\author{Jean L. Turner}
\institute{Jean L. Turner \at Department of Physics and Astronomy UCLA, Los Angeles CA 90095-1547 USA, \email{turner@astro.ucla.edu}}

\maketitle

\vskip -.5in
\noindent{\small ---contribution from {\it Astrophysics in the Next Decade: JWST 
and Concurrent Facilities}, Astrophysics and Space Science Proceedings, 2009, ed. H. A. Thronson, M. Stiavelli, and A. G. G. M. Tielens---}
\vskip .3in

\abstract*{Extreme star formation includes star formation in starbursts and
regions forming super star clusters. We survey the
current problems in our understanding of the star formation process in 
starbursts and super star clusters---initial mass functions, cluster mass functions,
star formation efficiencies, and radiative feedback into molecular clouds---that 
are critical to our understanding of the formation and survival of large star
clusters, topics that will be the drivers of the observations of the next decade.}

\abstract{Extreme star formation includes star formation in starbursts and
regions forming super star clusters. We survey the
current problems in our understanding of the star formation process in 
starbursts and super star clusters---initial mass functions, cluster mass functions,
star formation efficiencies, and radiative feedback into molecular clouds---that 
are critical to our understanding of the formation and survival of large star
clusters, topics that will be the drivers of the observations of the next decade.}

\section{Extreme Star Formation in the Local Universe}
\label{sec:1}

Extreme star formation is the violent, luminous star formation that occurs
in starbursts and luminous infrared galaxies. It is the formation of
super star clusters that may eventually become globular clusters. 
It is the source of galactic winds and metal enrichment in galaxies. It is probably
what most star formation in the universe was like several gigayears ago.

The process of star formation and its associated microphysics
 is most easily studied in the local universe  where we can examine
  the process of star formation in detail. While there
are regions in the Galaxy that may qualify as extreme star formation, 
most extreme systems are extragalactic. 
Advances in the study of extragalactic star formation during the next
decade are likely to come from
improvements in spatial resolution and sensitivity, particularly in the infrared
and submillimeter parts of the spectrum. The refurbished HST, forthcoming
JWST, and ground-based adaptive optics systems
 will make fundamental contributions to our understanding
of the stellar content of extreme star forming regions.
Herschel, ALMA, CARMA, Plateau de Bure, SMA, and SOFIA are
the far-infrared, submillimeter, and millimeter telescopes that
will deliver images and spectra of molecular gas in
galaxies, enabling the study of the earliest stages of star formation, and 
the regulation of star formation by feedback into molecular clouds.  
With the subarcsecond and milliarcsecond resolutions now
possible we can study the star formation process in other galaxies on the
parsec spatial scales of molecular cores, young clusters, and Stromgren spheres.

This review is an attempt to distill a very active area of research
on extreme star formation, covering both the stellar content and studies of the 
star-forming gas, and to project this research
into the observations of the next decade. The field is a remarkably broad
one, because in the process of star birth and cluster evolution, 
stars and gas are physically interrelated. The observations
discussed here cover the range from ultraviolet spectroscopy of
hot stars to millimeter line imaging of cold molecular clouds. The focus will be 
on star formation in the local universe where individual star-forming
regions can be resolved, and the star formation process itself can
be studied. Star formation in
the early universe, where extreme star formation may have been 
more the norm than the exception, is covered elsewhere in this volume
in contributions by Tom Abel and Alice Shapley. 

\section{What constitutes extreme star formation in the local universe?}
\label{sec:2}

Advances in instrumentation have shaped and refined our current view of  
``extreme star formation" (ESF). 
Many of the features of ESF in the local universe were known half a century ago: 
 giant H{\sc II} regions  \citep{1962ApJ...135..694B}, 
galactic winds \citep{1963AJ.....68R.284L, 1964AJ.....69R.535B}, young ``populous
clusters" \citep{1952PASP...64..196G, 1961ApJ...133..413H}, 
O-star dominated compact dwarf galaxies \citep{1970ApJ...162L.155S},
and  bright extragalactic radio 
sources \citep{1981ApJ...248..105W}.  The recognition
that there were individual star formation events that could energetically
dominate the evolution of a galaxy came with the development of infrared
and high resolution radio capabilities 
\citep{1975ApJ...197...17R, 1978ApJ...220L..37R, 1979ARA&A..17..477R, 
1983ApJ...267..551G}.
However, it was the IRAS all-sky survey in the mid and far-infrared 
that established the universality and energetic importance of the ``starburst" to galaxy evolution.  IRAS  demonstrated that the luminous output of galaxies can be 
dominated by infrared emission and recent star formation
\citep{1984ApJ...278L..67D, 1984ApJ...283L...1S, 1986ApJ...303L..41S, 
1987ARA&A..25..187S, 1987ApJ...320..238S}
and that extreme star formation may even be linked to the development of 
nuclear activity in galaxies \citep{1988ApJ...328L..35S}.  Many of the early IRAS
results have been followed up with the subsequent ISO \citep{2000ARA&A..38..761G}
and Spitzer \citep{2003PASP..115..897L} infrared space observatories.

The definition of starburst has evolved since the time of IRAS.
 In early incarnations it described systems
that would deplete their gas in substantially less than a Hubble time. However,  
this definition can exclude galaxies with substantial
reservoirs of gas that are forming stars at prodigious rates. 
Infrared luminosity can be used to classify these extreme star-forming systems:  
``Ultraluminous infrared galaxies" (ULIRGs) have luminosities 
of  $\rm L_{IR}>10^{12}~L_\odot$ 
\citep{1987ApJ...320..238S} and ``luminous" infrared galaxies have
$\rm L_{IR}>\sim 10^{11}~L_\odot$. These luminous galaxies 
owe most of their energetic output to star formation 
\citep{1986ApJ...305L..45S, 1998ApJ...498..579G}.
 Another definition captures the localized intensity of starbursts: 
 \citet{1998ARAA..36..189K}  defines starburst in terms of
a star-forming surface density, $100~\rm M_\odot~pc^{-2}~Gyr^{-1}$, or
in terms of luminosity, $10^{38.4}$--$10^{39.4}~\rm erg\, s^{-1}\, kpc^{-2}$. 

 \begin{figure}[ht]
\begin{center}
\includegraphics[scale=.2]{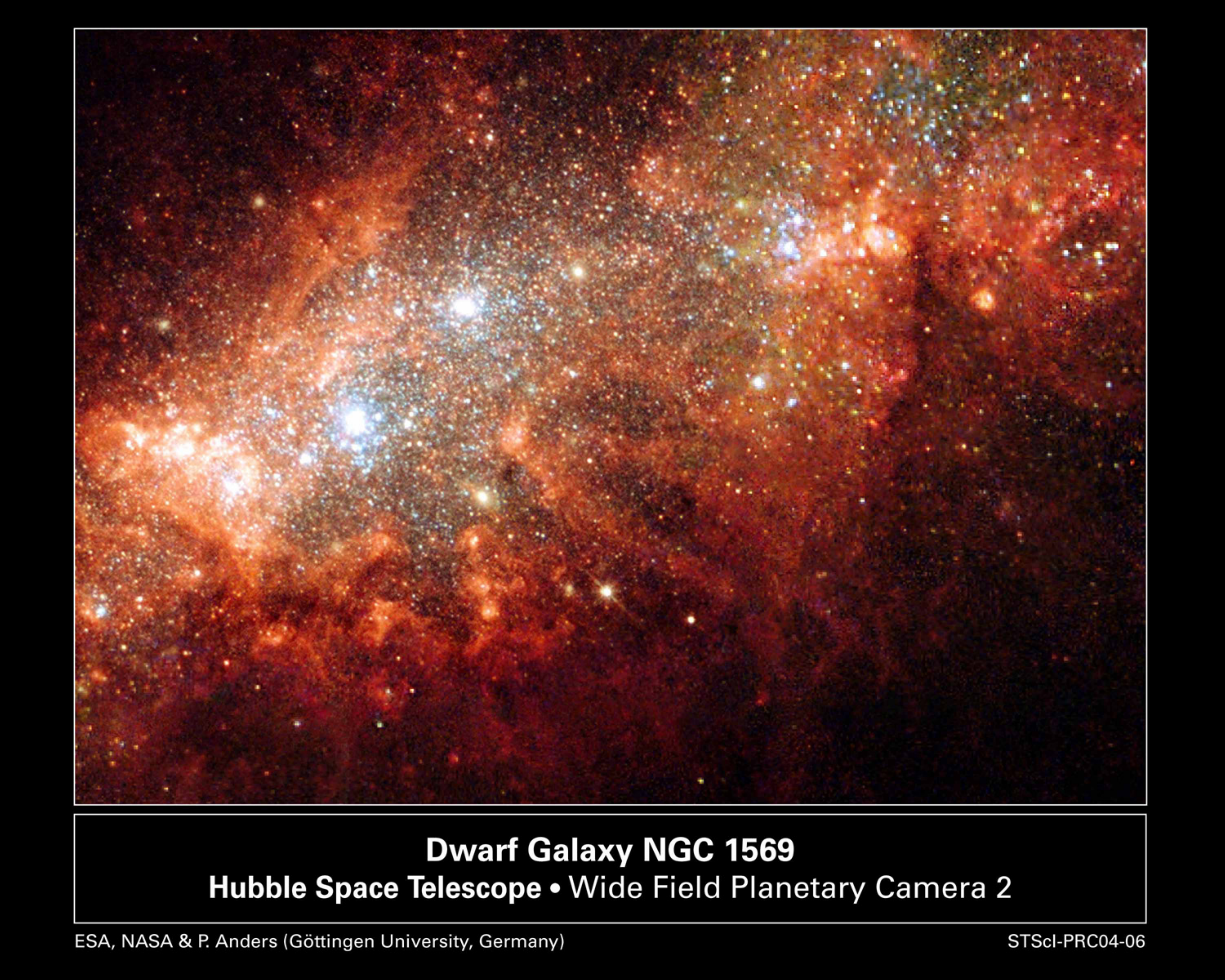}
\caption{HST revealed that the bright sources NGC 1569-A and NGC 1569-B were 
large clusters of stars. Credit: ESA, NASA, and P. Anders.}\label{fig:1}
\end{center}       
\end{figure}

High spatial resolution has also modified our view of starbursts,
revealing that they often, and perhaps nearly always,  consist of the formation of
large numbers of extremely large clusters, ``super star clusters." For the
purposes of this review, we will use the term super star cluster (SSC) to denote
massive young clusters of less than $\sim$100 Myr in age, and globular clusters
to be those systems more than 7--10 Gyr in age.
 Both starbursts and SSCs comprise ``extreme star formation."

The Hubble Space Telescope (HST) is largely responsible for the 
recognition of the ubiquity of SSCs and their importance in starbursts.
The idea that massive clusters similar to globular clusters
are actually forming in large numbers
at the present time  was slow to germinate, probably due to 
lack of spatial resolution and our inability to resolve them, 
although the possibility was recognized early on in the large
clusters of the Magellanic clouds \citep{1961ApJ...133..413H}. 
The cluster R136, the star cluster responsible for 
the lovely 30 Doradus Nebula in the Large Magellanic Clouds, was 
believed by many to be a single supermassive star before it was
 resolved with speckle observations from the ground \citep{1985A&A...150L..18W}.
Other mysterious objects included two bright sources in the nearby dwarf
galaxy NGC 1569, which were difficult to classify due to the uncertainty
in distance to this nearby galaxy.
Regarding the two ``super star cluster" candidates, Arp \& Sandage (1985) stated: 
\begin{quotation}
A definite resolution of the present problem in NGC~1569, and for 
the same problem with the bright object in NGC~1705, lies in the spatial resolution
 into stars of these three high-luminosity
blue objects using the imaging instrument of the wide-field camera of Space Telescope.
\end{quotation}
\nocite{1985AJ.....90.1163A}
HST did indeed resolve the objects in NGC~1569, revealing that they were large and luminous star clusters. In Figure \ref{fig:1}  
 is shown the HST image of NGC 1569,
with objects A and B referred to by Arp and Sandage.  Object A consists of two
superimposed clusters;  
crowded conditions and confusion complicate the study of SSCs, even 
in the closest galaxies and with the
angular resolution of HST!
R136 was resolved into a 
compact and rich cluster of stars 
by HST \citep{1993ApJ...419..658D,1996ApJ...459L..27H}. 
HST imaging also revealed super star clusters in NGC 1275, M82, NGC 1705, 
the Antennae, 
and numerous other starburst galaxies 
\citep{1992AJ....103..691H, 1993AJ....106.1354W,
1994ApJ...433...65O, 1995AJ....110.2665M, 1996AJ....111.2248M, 
1995AJ....109..960W}
The discovery of young, blue star clusters with luminosities 
consistent with those expected for young globular clusters
 in local starburst galaxies
 meant not only that conditions favorable to the formation of
 protoglobular clusters exist in the present universe, but also 
 that this extreme form of star formation
 is close enough for the star formation process itself to be studied.

\begin{table}                           
\caption{Super Star Clusters in Context}
\label{tab:1}      
\begin{center}
\begin{tabular}{lcccccc} 
\hline\noalign{\smallskip}
Type& $N_*$&Mass&$\rm r_h$&$\rho_h^a$&$\rm M_V$&Age\\
       &  & ($\rm M_\odot$) & (pc) & ($\rm M_\odot\, pc^{-3}$) & & (yrs) \\
\noalign{\smallskip}\svhline\noalign{\smallskip}
globular cluster &$>10^5$&$10^{3.5}$--$10^6$&0.3--4&$10^{-1}$--$10^{4.5}$&-3 to -10& $>10^{10}$\\
open cluster    & 20--2000& 350--7000     & 2.5--4.5& 1--100 & -4.5 to -10 
&$10^6$--$10^{9.8}$\\
embedded cluster& 35--2000 & 350--1100 & 0.3--1 & 1--5 & \dots & $10^6$--$10^{7}$\\
SSC & $>10^5$  & $10^5$--$10^6$ & 3--5 & \dots  & -11 to -14 & $10^6$--$10^{7}$\\
\noalign{\smallskip}\hline\noalign{\smallskip}
\end{tabular}
\end{center}
$^a$Half mass mean density. Number of members, $N_*$, is not as well-defined
for the larger clusters as it is for the local open and embedded clusters.
References. Battinelli \& Capuzzo-Dolcetta 1991. Billett et al.\ 2002. Harris 1996.
Harris \& Harris 2000.  Lada \& Lada 2003. Mackey \& Gilmore 2003. 
McLaughlin \& Fall 2008. McLaughlin \& van der Marel.
2005. Noyola \& Gebhardt 2007. 
\end{table}

\nocite{1991MNRAS.249...76B}  
\nocite{2002AJ....123.1454B}  
\nocite{1996AJ....112.1487H}  
\nocite{2000Harrissquare} 
\nocite{2003ARA&A..41...57L} 
\nocite{2003MNRAS.338...85M}  
\nocite{2008ApJ...679.1272M}  
\nocite{2005ApJS..161..304M}  
\nocite{2007AJ....134..912N}
Super star clusters appear to be sufficiently massive 
and rich to be globular clusters, differing from them only in age.
Table \ref{tab:1} lists the general characteristics of different classes of 
Galactic star clusters and SSCs. 
The brightest young SSCs have $M_V \sim -14$. They are brighter than 
globular clusters because of their youth. 
It is convenient to take the lower bound for SSCs to be $M_V\sim -11$
\citep{2002AJ....123.1454B}, approximately 
the magnitude of R136 in the LMC, which is also comparable in size to the 
most massive young Galactic clusters. However, R136 is considered by some to 
be on the small side for a globular cluster. The older LMC cluster 
NGC~1866, at $\rm L \sim 10^6~\rm L_\odot$ and 
an intermediate age of 100 Myr,   is closer to a 
genuine globular cluster \citep{1993ASPC...48..588M}. The upper limit to
the ages of SSCs is also somewhat arbitrary; while there is evidence that 
typical cluster dissolution timescales are about 10 Myr, there are also
intermediate age clusters with ages of $\sim$1~Gyr even within the Local Group.

\section{Extreme Star-forming Regions of the Local Universe}
\label{sec:3}

The most luminous young SSCs of the Galactic neighborhood are the testbeds for
study of the star formation process in large clusters and in starburst systems. What are the Orions of the SSC world? In Table~\ref{tab:2} we have compiled from
the literature properties of some well-studied and
spatially resolved SSCs in the local universe. 
Included are Galactic center clusters and large Galactic star-forming regions.
While smaller than many extragalactic SSCs, these Galactic clusters are close and 
more easily studied and should share many of the star-forming properties.
The super star clusters of Table~\ref{tab:2} reflect a wide range of environments and 
evolutionary stages in the formation and evolution of SSCs. 

\begin{table}
\caption{Massive Young Star Clusters in the Local Universe}
\label{tab:2}       
%
%
\begin{tabular}{llcccccccc}
\hline\noalign{\smallskip}
Host&Cluster&D&R$^a$&log $\rm {{L_*}\over{L_\odot}}$ &log $\rm {{M_*}\over{M_\odot}}$
&log $\rm N_{Lyc}$ & $\rm N_O$ & $\rm M_V$ & Age \\
&&(Mpc) & (pc)   &  &&& &&(Myr) \\
\noalign{\smallskip}\svhline\noalign{\smallskip}
Galaxy & Arches           &0.008   & $>$0.5   & 8.0   & 4.1        & 51.0  &  160  & \dots &  2-2.5 \\
Galaxy & Quintuplet     &0.008   & 1.0    &  7.5  & 3-3.8 & 50.9  &   100   & \dots &  3-6  \\
Galaxy & Center          &  0.008  & 0.23  & 7.3    & 3-4 & 50.5  & 100     &  \dots & 3-7 \\
Galaxy & Sgr B2       & 0.008  & 0.8  &  7.2    & \dots      & 50.3  & (100)  & \dots  &\dots  \\ 
Galaxy & NGC 3603     &0.0076  & 4.5   & 7.0     & 3.4    & 50.1   &$>$50 & \dots &1-4  \\   
Galaxy & Westerlund 1 & 0.0045 &1     &  \dots &  4.7    &  51.3   & 120  &  \dots &  3-4   \\
Galaxy & W49A           &0.014 &  5     & 7.4     & \dots     & 50.1    & 80    & \dots  & \dots \\
LMC &    R136$^c$      & 0.05  & 2.6   & \dots  &  4.8       & 51.7     & $>$65& -11  &1-3  \\
NGC1569&NGC1569-A1&2.2&   1.6-1.8 &  \dots &     6.11    &   \dots    &    \dots  &-13.6$^b$ &    \dots \\
NGC1569&NGC1569-A2&2.2&   1.6-1.8 &\dots  &    5.53   & \dots & \dots& \dots &\dots \\
NGC1569&NGC1569-B &  2.2  &   3.1  &\dots   & 5.6   &  \dots  & \dots & -12.7 & 15-25  \\
NGC 1705&NGC1705-1&  5.3  &  1.6     & \dots  & 5.68  & $<$51  & \dots  & -14.0 & 12 \\
He 2--10  &He 2-10-1   &   3.8 &  1.5  &  \dots &    5.7  &  \dots  &  1300   &  -14.3  & 5.2\\
He 2--10   &He 2-10-A-4   &  3.8  & 3.9 &   \dots & \dots & 52.4 & \dots & \dots &\dots \\
He 2--10  & He 2-10-A-5  & 3.8    & 1.7  &\dots   & \dots &  51.9 & \dots & \dots & \dots \\
He 2--10  & He 2-10-B-1  & 3.8    & 1.8  &\dots   & \dots &  51.9 & \dots & \dots & \dots \\
He 2--10  & He 2-10-B-2  & 3.8    & 1.8  &\dots   & \dots &  52.0 & \dots & \dots & \dots \\
M82 & M82-A1	      & 3.6     &  3.0  &   7.9   &    6.0    &  50.9    &  100  & -14.8    & 6.4 \\
M82 	& M82-F         &  3.6   &  2.8   &   7.73  &  5.8    &  \dots  & \dots &  -14.5 & 50--60 \\
M82 & M82-L          & 3.6   &  \dots &  \dots & 7.6      &   \dots    &  \dots &  \dots & \dots \\
NGC3125&NGC3125-A1&11.5 & \dots & \dots&\dots&52.4$^d$&250--3000$^e$& \dots&3--4 \\
NGC3125&NGC3125-A2& 11.5 & \dots &\dots &\dots &\dots& 550--3000 & \dots & 3--4 \\
NGC3125&NGC3125-B1,2 &11.5 & \dots &\dots &\dots& 52.2 &  450    &   \dots   & 3--4 \\
Antennae&Antennae-IR& 13.3   & $<$32  &  \dots  & 6.48& 52.6 & 120 &  -17$^f$ &  4\\
NGC4214&NGC4214-1&  4.1 &  $<$2.5&\dots &\dots & \dots  &  280   &  -13.1  & \dots \\
NGC5253&NGC5253-5  & 3.8 & \dots&  5.8 & \dots & 51.9 & 155    & $\sim$-14    & 2 \\
NGC5253&NGC5253-IR & 3.8  & 0.7 &  9.0 & \dots &52.5&1200-6000$^g$&\dots &2--3 \\

\noalign{\smallskip}\hline\noalign{\smallskip}
\end{tabular}
 $^a$Cluster radii are half-light radii. 
$^b$Cluster A is two clusters, de Marchi et al.\ 1997. 
$^c$Bright core of a larger, complex cluster, NGC 2070.
$^d$A1 and A2.
$^e$Range in O stars is due to differences in reddening. 
$^f \rm  M_K.$ 
$^g$Resolved source; lesser number for $r<$1 pc. 
References. {Arches, Quintuplet, Galactic nuclear center clusters:}  
Figer et al.\ 1999, Lang et al.\ 2001, Figer et al.\ 2005,
Stolte 2003, Stolte et al. 2002, 2005, 2007, Najarro et al.\ 2004,
Figer 2003, 2004, 2008. Kim et al. 2000, 2004, 2007, Kim \& Morris 2003.
{Sgr B2:} Dowell 1997, Gaume et al.\ 1995.
{NGC 3603:} de Pree, Nysewander, \& Goss 1999, Eisenhauer et al.\ 1998, 
Pandey et al.\ 2000, Drissen et al.\ 2002, N{\"u}rnberger \& Petr-Gotzens 2002,
Stolte et al.\ 2006, Harayama et al.\ 2008, Melena et al.\ 2008.
{W49:} Smith et al.\ 1978, Welch et al.\ 1987, Conti \& Blum 2002, Homeier \& Alves 2005.
{Westerlund 1:} Clark et al.\ 1998, 2005, N{\"u}rnberger et al.\ 2002, N{\"u}rnberger 2004, 
Crowther et al.\ 2006, Mengel \& Tacconi-Garman 2007, 2008, Brandner et al.\ 2008.
{R136:} Mills et al.\ 1978, Meylan 1993, Hunter et al.\ 1995, Massey \& Hunter 1998,
Noyola \& Gebhardt 2007.
{M82-A1:} Smith et al.\ 2006. 
{M82-F:} Smith \& Gallagher 2001, O'Connell et al.\ 1995, McCrady et al.\ 2005.
{M82-L:} McCrady \& Graham 2007.
{NGC1569 A and B:} O'Connell et al.\ 1994, Sternberg 1998,
Hunter et al.\ 2000, Ho \& Filippenko 1996a, Smith \& Gallagher 2001, Origlia et al.\ 2001, 
Gilbert 2002, Larsen et al.\ 2008.
{NGC1705-1:} Ho \& Filippenko 1996b, Sternberg 1998,
Smith \& Gallagher 2001, Johnson et al.\ 2003, V{\'a}zquez et al.\ 2004.
{He~2-10-1:} Chandar et al.\ 2000. One of five clusters within He 2-10A.
{He~2-10-A, B:}  Vacca \& Conti 1992, Johnson \& Kobulnicky 2003.
{NGC3125:} Vacca \& Conti 1992, Schaerer et al.\ 1999ab, Stevens et al.\ 2002, 
Chandar et al.\ 2004, Hadfield \& Crowther 2006. 
Region A has log$Q_0 = 52.39$, for 4000 O stars, region
B  log$Q_0 = 52.19$, for 3200 O stars, Hadfield \& Crowther. 
{Antennae:} Gilbert et al.\ 2000, for 13.3~Mpc. 
{NGC5253-5:} Gorjian 1996, Calzetti et al.\ 1997, Schaerer et al.\ 1997,
Tremonti et al.\ 2001, Chandar et al.\ 2004, Vanzi \& Sauvage 2004, Cresci et al.\ 2005.
{NGC5253-IR:} Obscured IR/radio source offset by 
$\sim$0.5$^{\prime\prime}$ from NGC5253-5. 
Beck et al.\ 1996, Turner et al.\ 1998, 2000, 2003, Mohan et al.\ 2001,
Alonso-Herrero et al.\ 2004, Turner \& Beck 2004, 
Mart{\'i}n-Hernandez et al.\ 2005, Rodr{\'i}guez-Rico et al.\ 2007.
\end{table}

\nocite{1999ApJ...525..750F} \nocite{2003ANS...324..255F} 
\nocite{2004ASPC..322...49F} \nocite{2005Natur.434..192F}
 \nocite{2008arXiv0803.1619F}  \nocite{2008IAUS..250..247F} 
\nocite{2001AJ....121.2681L} \nocite{2003PhDT.........2S}  
\nocite{2002A&A...394..459S} \nocite{2005ApJ...628L.113S} 
\nocite{1998Ap&SS.263..251N} 
\nocite{2000ApJ...545..301K}  \nocite{2004ApJ...607L.123K} 
 \nocite{2007JKAS...40..153K}
\nocite{1995ApJ...449..663G} \nocite{1997ApJ...487..237D}
 \nocite{1998ApJ...498..278E} \nocite{1995RMxAA..31...39D} 
\nocite{2000PASJ...52..847P} \nocite{1995AJ....110.2235D} 
\nocite{2002A&A...382..537N} \nocite{2006AJ....132..253S} 
\nocite{2008ApJ...675.1319H} \nocite{2008AJ....135..878M}
\nocite{1999AJ....117.2902D}
\nocite{1978A&A....66...65S} \nocite{1987Sci...238.1550W} 
\nocite{2002ApJ...564..827C} \nocite{2005A&A...430..481H}
\nocite{1998MNRAS.299L..43C} \nocite{2005A&A...434..949C} 
\nocite{2002A&A...394..253N} \nocite{2004PhDT.........8N} 
\nocite{2006MNRAS.372.1407C} \nocite{ 2008A&A...478..137B} 
\nocite{2007A&A...466..151M} \nocite{2008arXiv0803.4471M}
\nocite{2008A&A...478..137B}
\nocite{1978MNRAS.185..263M} \nocite{1995ApJ...448..179H} 
\nocite{1998ApJ...493..180M} \nocite{2000ApJ...533..203S}
\nocite{1993ASPC...48..588M}\nocite{2007AJ....134..912N}
\nocite{2006MNRAS.370..513S} \nocite{1995ApJ...446L...1O}
 \nocite{2001MNRAS.326.1027S} \nocite{2003ApJ...596..240M}
\nocite{2007ApJ...663..844M}
\nocite{1994ApJ...433...65O} \nocite{1996ApJ...466L..83H} 
\nocite{1998ApJ...506..721S} \nocite{2000AJ....120.2383H} 
\nocite{2001AJ....122..815O} \nocite{2001MNRAS.326.1027S} 
\nocite{2002PhDT........15G} \nocite{2008MNRAS.383..263L}
\nocite{1996ApJ...472..600H} \nocite{1998ApJ...506..721S} 
\nocite{2001MNRAS.326.1027S} \nocite{2003AJ....126..101J} 
\nocite{2004ApJ...600..162V}
\nocite{2003ApJ...586..939C} \nocite{1992ApJ...401..543V} 
\nocite{2003ApJ...597..923J} 
\nocite{1992ApJ...401..543V} \nocite{1999A&A...341..399S} 
\nocite{1999A&AS..136...35S} \nocite{2002MNRAS.335.1079S} 
\nocite{2004ApJ...604..153C} \nocite{2006MNRAS.368.1822H}
\nocite{2002PhDT........15G} \nocite{2000ApJ...533L..57G} 
\nocite{1996AJ....112.1886G} \nocite{1997AJ....114.1834C} 
\nocite{1997ApJ...481L..75S} \nocite{2001ApJ...555..322T} 
\nocite{2004A&A...415..509V} \nocite{2005A&A...433..447C}
\nocite{1996ApJ...457..610B} \nocite{1998AJ....116.1212T}
\nocite{2000ApJ...532L.109T} \nocite{2003Natur.423..621T} 
\nocite{2001ApJ...557..659M} \nocite{2004ApJ...612..222A} 
\nocite{2004ApJ...602L..85T} \nocite{2005A&A...429..449M} 
\nocite{2007ApJ...670..295R}

The Galactic massive young clusters are 
readily resolved into stars and contain a wealth of information
on young massive cluster evolution. However, even for these nearby clusters, 
confusion, contamination, and rapid dynamical evolution
introduce great complexity
into observational interpretations. Westerlund 1 is the closest
of these large clusters, located in the Carina arm.  NGC 3603
is a large southern cluster somewhat more distant.  The Arches, Quintuplet, and
 Galactic Center nuclear clusters   have formed 
  in the  immediate vicinity of a supermassive black hole, and may differ
  in structure and evolution from large clusters in more benign environments. 
The study of the Galactic Center clusters has been made possible by
high resolution infrared observations. Included in Table \ref{tab:2} are luminous
embedded star-forming regions SgrB2 and W49A; their relation to
the massive, unembedded star clusters is unclear, although they have similar
total luminosities. The other SSCs
listed are in galaxies within $\sim$20 Mpc, in which clusters
can be spatially resolved. Many of these SSCs have been identified by 
their location within ``Wolf-Rayet" galaxies,
those galaxies with a strong He{\sc II}~4686 line indicating the
presence of significant numbers of Wolf-Rayet stars of age $\sim$3--4~Myr 
\citep{1991ApJ...377..115C,1999A&AS..136...35S}.  The Wolf-Rayet feature
is a relatively easy way to identify in systems with large clusters of young
stars, and these are often found in SSCs.
Many of the clusters in Table~\ref{tab:2} have dwarf galaxy hosts. 
This may be a selection effect
due to the difficulty of isolating clusters amid the higher confusion and extinctions
in large spirals, since SSCs are definitely present in many local spirals, such as 
NGC 253, NGC 6946, and Maffei 2 
\citep{1992ARA&A..30..575C, 1994ApJ...421..122T, 1996AJ....112..534W, 
1996ApJS..107..215M, 2001AJ....121.3048M, 
2006ApJ...644..914R, 2006AJ....132.2383T, 2008A&A...483...79R}.
The dominance of dwarf galaxy hosts
may also be due to ``downsizing," the tendency for star 
formation to occur in smaller systems at later times. 

It is evident from Table~\ref{tab:2} that it can be difficult to compare these 
clusters because embedded and visible clusters are characterized in different ways. 
Embedded
clusters are often characterized by photons that have been absorbed by gas or 
dust, with well-defined Lyman continuum fluxes and infrared luminosities. 
Visible clusters have star counts, cluster magnitudes, colors, and stellar velocity dispersions;
these clusters can have good masses and ages. Putting together an evolutionary
sequence of objects thus requires multiwavelength observations at high spectral
resolution. Extinction is observed to decrease with increasing cluster age 
\citep{2005A&A...443...41M}, 
as one might expect  from Galactic star-forming regions, 
so the embedded clusters are likely to also be the youngest clusters.

High angular resolution is key to the study of even the closest super star clusters,
which are often found forming in large numbers.
One of the first known super star clusters, NGC 1569-A, consists of two
superimposed clusters \citep{1997ApJ...479L..27D}, which is not immediately
obvious even in the HST image (Figure~\ref{fig:1}). 
The embedded IR/radio SSC
in the center of NGC~5253 was found to be offset by a fraction of an arcsecond
from the brightest optical cluster, NGC 5253-5, \citep{1997AJ....114.1834C}
 only 5--10 pc away \citep{2003Natur.423..621T, 2004ApJ...612..222A}.  

Extinction can also be extreme in bright IR-identified starburst regions,
and can obscure even the brightest clusters through the near-IR.  Observed
differential extinctions between the infrared Brackett lines at 2 and 4$\mu$m
indicate $A_V > 1$ and even $A_K >1$ in many starbursts 
\citep{1989ApJ...337..230K, 1990ApJ...349...57H}. IR-derived extinctions
are often higher than those derived from Balmer recombination
lines toward the same regions \citep{1979ApJ...227...64S} because of
extinctions internal to the H{\sc II} regions themselves.
In M82, near-IR and mid-infrared spectroscopy indicates
extinctions of $A_V\sim 25$ 
\citep{1977ApJ...217L.121W,1979ApJ...227...64S} to 
$A_v\sim 50$ \citep{2001ApJ...552..544F}, similar to values
observed in Galactic compact H{\sc II} regions, but over much larger areas.
The clusters in Arp 220 are heavily obscured, with estimated $A_V \sim 10$--45~mag
\citep{2001MNRAS.321...11S, 1998ApJ...498..579G}; regions behind the
molecular clouds can reach $A_V \sim 1000$ \citep{1998ApJ...507..615D}.

What does the next decade hold for 
 the clusters of Table \ref{tab:2} and other nearby clusters like them?
First, there are more SSCs to discover in the local universe, particularly
embedded ones.  IRAS is still a valuable tool for discovering young SSCs, but
with arcminute resolution, it is not sensitive to bright, subarcsecond sources.
There undoubtedly remain many compact, young ESF events
to be found in the local universe. The WISE mission, an all sky mid-IR survey, will provide an
extremely valuable dataset for discovering young and embedded SSCs. 
The enhanced sensitivities and high spatial resolution of the next
generation of telescopes (JWST, EVLA, ALMA, SOFIA), will redefine our concept of
``local" SSC formation, extending
 this list to more distant systems and to young SSCs within large, gas-rich
spirals. The near-infrared, in particular, is an valuable link between visible and 
embedded SSCs. Subarcsecond resolutions are necessary to resolve 
individual clusters in regions of high and patchy extinction, and the closest
galaxies are even now being pursued with adaptive optics.
JWST and future extremely large ground-based telescopes
will play an important role in connecting embedded clusters to their older, visible
siblings to enable a longitudinal study of SSC evolution.

\section{Initial Mass Functions and the Most Massive Stars in SSCs}
\label{sec:4}

The initial mass functions (IMFs) of SSCs have important consequences for
cluster masses, their long-term survival, and potentially, for the eventual remnants
left by their dissolution. 
Are the IMFs of clusters power law? Is there evidence for top-heavy
IMFs? How do the IMFs of SSCs compare
to Galactic IMFs?  Is there evidence for initial mass segregation in 
young clusters? The most massive star in the universe is likely to be in an SSC; is there
a fundamental limit to the mass of stars?  There is a
 review of the outpouring of recent IMF
work on massive young clusters  by Elmegreen (2008). 

Mass functions  have been determined for the large star clusters of the Local Group.
Many appear to be Salpeter. The Salpeter mass functions are defined as 
$\rm \xi(M) \sim M^{-\alpha}$, where $\alpha = 2.35$, or, expressed 
 in logarithmic mass intervals as
$\rm \xi_L\, d \log M \sim M^\Gamma d \log M$, with $\Gamma =-1.35$ 
\citep{1986FCPh...11....1S, 1998ASPC..142..201S}. The
Kroupa IMF is Salpeter at higher masses, and
 flattens to $\alpha = 1.3$ for stars below 0.5~{M$_\odot$} \citep{2001MNRAS.322..231K}. 
 Kroupa IMFs often cannot be distinguished from Salpeter in
 extragalactic SSCs. We will
 adopt the $\Gamma$ convention here and note that while in many cases these are referred to
 as IMFs, what is observed is actually a present day mass function (PDMF), from which
 an IMF may be inferred.
 
The PDMF of R136 in the LMC
 has been measured down to 0.6~M$_\odot$ \citep{1998ApJ...493..180M}, 
 where it is consistent with Salpeter,  $\Gamma \sim {-1.3}$,
 and then  flattens below  2~M$_\odot$ to $\Gamma \sim -0.3$ \citep{2000ApJ...533..203S}. 
For a Salpeter or Kroupa power law IMF, flattening or turnover of the power law
 corresponds to an effective  ``characteristic mass" for
the cluster \citep{2003ARA&A..41...57L}.  R136 thus appears to have a
characteristic mass of 1--2~M$_\odot$, as compared to $\sim$0.5--1~M$_\odot$ for the much
smaller nearby Galactic embedded clusters \citep{2003ARA&A..41...57L}. 
The southern cluster NGC 3603 has a somewhat flatter-than-Salpeter power
law slope of $\Gamma=-0.7$ to $ -0.9$ from 0.4 to 20~M$_\odot$ 
 \citep{2004AJ....127.1014S, 2006AJ....132..253S,2008ApJ...675.1319H}.
Arches has a very similar power law PDMF
with $\Gamma = -0.8$ down to 1.3~M$_\odot$ \citep{2002A&A...394..459S}, 
which may correspond
to an IMF that is close to Salpeter $\Gamma = -1.0$--1.1 \citep{2007JKAS...40..153K}. 
The Arches cluster is mass segregrated, with   
a flatter slope in the center, $\Gamma \sim 0$, than in the outer parts of the cluster,
 and  the mass function may turn over at 6--7~M$_\odot$ in the core
 \citep{2005ApJ...628L.113S}; however,
the MF at larger radii does not show this effect \citep{2007JKAS...40..153K}.  
Trends appear to be toward flatter MF power laws and higher characteristic
masses for the largest clusters in the Local Group, but the numbers of clusters
are very small, and dynamical effects are poorly understood as yet 
\citep{2002A&A...394..459S}.

Beyond the Local Group, it is more difficult to determine IMFs. 
Observational quantities for more distant SSCs are 
integral properties such as total luminosity and dynamical masses from cluster
velocity dispersions.  Constraints on IMFs  from these integral properties
require assumptions about mass cutoffs, cluster ages, and cluster structure. 
There is accumulating evidence, however, that IMFs in starbursts and the IMFs
in SSCs, if Salpeter, may have higher characteristic cutoffs than typical
Galactic clusters. Rieke et al.\ (1980) \nocite{1980ApJ...238...24R} 
modeled the starburst in M82 from its IR emission properties, and 
concluded that the IMF of the starburst 
must have a low mass cutoff of 3--8~M$_\odot$.  Sternberg (1998) used 
visible mass-to-light ratios to argue for a cutoff of 1~M$_\odot$ for NGC1705-1,
although he does not find a cutoff for NGC~1569A. If the IMF in the M82
SSCs is Salpeter, then McCrady et al.\ (2003) \nocite{2003ApJ...596..240M}
find that the individual M82 clusters must have truncated mass functions, although 
some clusters show 
strong evidence for mass segregation and possible dynamical evolution, which
complicates this interpretation 
\citep{2003ApJ...596..314M, 2005ApJ...620L..27B, 2005ApJ...621..278M, 2007ApJ...663..844M}.

What the low mass cutoffs are for SSCs, whether this varies with environment
and how, and what is the likely cause of the low mass cutoffs, effective
characteristic masses,  and the 
effects of mass segregation on cluster masses and evolution will be 
fertile ground for SSC research in the next decade.

At the other end of the IMF, there is the question of what is the limiting mass of a star.
If the cluster IMFs bear any resemblance to the Galactic 
power law Salpeter function, then newly-formed clusters consisting
of hundreds of thousands to millions of stars are the place to find the most massive
and rare
O stars.  In addition, in dense star clusters there is  the possibility of the formation
of very massive stars through stellar collisions, which becomes a viable mechanism
in dense environments \citep{1998MNRAS.298...93B,2008arXiv0805.1176B}. 
What is the limiting mass of stars in the local universe? Where are the most
massive stars found?

Spectroscopy of individual stars is the most reliable way to identify the most
massive stars, but is only possible within the Local Group. 
Ultraviolet spectroscopy using the STIS instrument
on HST has allowed the classification of 45 known 
stars  of spectral class O2 and O3;  of these, 35 are found in the LMC
cluster R136 \citep{2002AJ....123.2754W}. R136 alone contains more than
65 O stars \citep{1998ApJ...493..180M}.  In addition to these visible O stars,
there is also a significant population
of infrared O stars and Wolf-Rayet stars in the Galactic Center, including
the three large star clusters, Arches, Quintuplet, and Galactic Center. These
infrared clusters, each about an order of magnitude less massive than R136,
 contain an estimated 360 O stars among them, nearly 60
Wolf-Rayet stars, and 2-3 LBVs  \citep{2008IAUS..250..247F}. 

The highest inferred stellar mass in the Local Group is $\sim 170~\rm M_\odot$ 
in the LMC, and $\sim 200~\rm M_\odot$ in the Galaxy; however the
latter could eventually turn out to be a binary system; the largest dynamical
mass measured is 90~$\rm M_\odot$ \citep{2002AJ....123.2754W}.
Weidner \& Kroupa (2004) \nocite{2004MNRAS.348..187W}  
estimate that based on its IMF, observed to be Salpeter, there should be one
750~M$_\odot$ star in the R136 cluster, and
 instead the upper mass limit appears to 
be 200~M$_\odot$ \citep{2006MNRAS.365..590K}. 
Oey \& Clarke (2005) extend this to a larger sample of OB associations,
\nocite{2005ApJ...620L..43O}
obtaining a cutoff of 120--200~M$_\odot$, although they caution that this cutoff
is sensitive to the IMF slope for stars $>$10~M$_\odot$.
Figer (2005) \nocite{2005Natur.434..192F} 
finds that there should be one star of M$\sim$ 500~M$_\odot$ 
in the Galactic center, where the
current upper mass limit instead seems to be 130~M$_\odot$. 
These observations argue for a stellar upper limit
close to the observed 150--200~M$_\odot$.  However, given the small
statistics, it may be that the absence of extremely
massive stars is simply an evolutionary effect \citep{2008arXiv0803.3154E}.  
For example, modeling suggests that the Pistol Star in the Galactic Center may have had an
initial mass of 200-250~M$_\odot$ \citep{1998Ap&SS.263..251N, 1998ApJ...506..384F}.
Stars with masses of several hundred times solar would evolve rapidly
\citep{2008A&A...477..223Y}, will lose significant fractions of their initial masses,
and in addition, might be dust-enshrouded for most of their lives. 
Moreover small number statistics at the upper mass end mean that for most
clusters, the power law slopes of IMFs are uncertain to a few tenths for clusters of
the size of NGC~3603 \citep{2008arXiv0803.3154E}.
Supermassive stars are elusive by nature.

If extremely high mass stars do exist and we have not seen them simply because of
their rapid evolution,  the best place to look for them is in the youngest
and largest clusters. These will probably be found in starbursts. There is a good chance
that the youngest regions containing the most massive stars will be deeply embedded, 
perhaps extinguished even in the near-infrared. 
In these cases, nebular diagnostics provide another way of gauging the upper end 
of the mass function.
The mid-infrared has a number of fine structure lines that can be observed in
large H{\sc II} regions in external galaxies
 \citep{2000ApJ...532L..21H, 2000ApJ...539..641T, 2006ApJ...646..161D}, including
fine structure lines of Ar, Ne, S, and O, that can provide valuable nebular diagnostics
of the input stellar radiation field even in embedded sources.  
 
Line ratios of the mid-IR nebular fine structure lines in starburst galaxies measured with the 
SWS spectrometer on ISO revealed unexpectedly low excitation H{\sc II} regions.
Nebular models of the ISO lines  indicate upper mass cutoffs of 
 $\rm M_{upper} \sim 30$~M$_\odot$ 
 for the IMFs in these starbursts \citep{2000ApJ...539..641T}.  
Given the luminosities and inferred stellar masses, the low
upper mass cutoffs are surprising. However, there are a number of possible
explanations for low excitation that would allow for the presence
of more massive stars \citep{2004ApJ...606..237R}. One of these is lack of
spatial resolution. Starburst regions often have extended regions of
ionized gas which, when combined with the comparatively 
hard spectra of the compact SSC H{\sc II} regions, will tend to wash out the high excitation 
lines. That the lines of [O{\sc IV}]25.9$\mu$m, 
[Ne{\sc III}]15.55$\mu$m, and [S{\sc IV}]10.51$\mu$m have been detected in dwarf 
starbursts suggests that hot stars are indeed present in some SSCs 
\citep{1996ApJ...457..610B, 1998A&A...333L..75L, 1999MNRAS.304..654C,
2006MNRAS.372.1407C, 2007AJ....134.1237B}.
Strong dependences on metallicity and differences in the input radiation
fields from existing stellar models that must be considered in these
interpretations \citep{1999MNRAS.304..654C, 2002A&A...389..286M, 2004ApJ...606..237R}.

High spatial resolution and spectroscopy from space will allow great improvements in our 
knowledge of the most massive stars and 
the upper mass cutoffs of SSC IMFs in external galaxies in the next decade.  Optical
and ultraviolet emission lines including Wolf-Rayet features will continue to be important
diagnostics of the high mass stellar content and ages of SSCs in an enlarged
sample of sources.  Access to mid-IR fine structure lines from space, via JWST,
 will give valuable information on the most massive stars in embedded young SSC nebulae,
 allowing observation of homonuclear line ratios of Ar and Ne that are not possible 
 from the ground. The improvement in spatial resolution is also extremely important
 for isolating the spectral signatures of compact SSC nebulae. 
The MIRI IFU on JWST, with order-of-magnitude improvements in sensitivity and
spatial resolution over previous instruments, will be an extremely powerful tool
for isolating compact SSC
nebulae from more diffuse and potentially lower excitation ionized gas within
the galaxies. The form of the cluster IMFs is key input to the question of the
long term survivability of clusters.

\section{What are the initial cluster mass functions for SSCs?}
\label{sec:5}

How are SSCs related to globular clusters? Are young SSCs
in the local universe precursors to globular clusters? One characteristic that
could link globular clusters to SSCs is the cluster mass function.
Globular clusters in the Galaxy have a characteristic mass of a few 
$\times 10^5$~M$_\odot$,
reflecting what appears to be a near-universal globular cluster mass function 
(GCMF) \citep{1998ASPC..136...33H}. 
Were globular clusters born with this mass function? Or is the present globular cluster
mass function the result of evolution due to dynamical forces such
as tidal stripping and disk shocking over billions of years
 \citep{1977MNRAS.181P..37F, 1997ApJ...474..223G, 2007arXiv0711.3540D, 
  2007MNRAS.377..352P, 2008ApJ...679.1272M}? 
SSCs, as potential precursors to globular clusters, could give us valuable
information about the nature vs. nurture question for the GCMF. Is the
initial cluster mass function (ICMF) for SSCs universal, and if so, what is it?
How does it evolve?

The ICMF has been extensively studied in the Antennae galaxies, where large
numbers of SSCs facilitate the statistics.
The Antennae are  a nearby (13.3 Mpc, Saviane et al.\ 2008) 
\nocite{2008ApJ...678..179S} IR-bright pair of gas-rich interacting galaxies. HST
 images revealed a system of  thousands of young SSCs spread across the
 galaxy pair  
 \citep{1995AJ....109..960W, 1999AJ....118.1551W}. ISO imaging 
 determined that the youngest  clusters and brightest infrared emission occur in a dusty and embedded region
 between the two galaxies \citep{1998A&A...333L...1M}.
Cluster masses and ages for the visible SSCs of the Antennae 
have been obtained from multicolor (UBVRI and H$\alpha$) HST photometry  
combined with modeling of the cluster colors and luminosities with
\textsc{STARBURST99} \citep{1999ApJS..123....3L, 2005ApJ...621..695V}
The cluster mass function (CMF) in the Antennae
 is power law with $\alpha \sim -2$ over
the range of masses $10^4$--$10^6$~M$_\odot$ \citep{2005ApJ...631L.133F}. 
The median cluster mass in the Antennae appears to be about an order of magnitude
less than the median globular cluster
mass of giant ellipticals such as M87 \citep{2002IAUS..207..545H}.
Ages determined from the SSC colors in the Antennae
range from $10^6$--$10^8$ yr \citep{1999ApJ...527L..81Z}, with a median
of 10-20 Myr \citep{2005ApJ...631L.133F, 2005A&A...443...41M}, but with a
small population extending to more than $10^{9}$~yrs. 
If the SSC system of the Antennae is typical, 
one could infer that the characteristic mass of globular clusters is due
to selective evolution of clusters of different mass \citep{2001ApJ...561..751F}. 
However, the fact that 70\% of the clusters in the Antennae
are less than $\sim$10--20 Myr of age could indicate that 
very few if any of the Antennae clusters will survive to become globular clusters.

Cluster mass functions have been determined for a number of other nearby
galaxies, mostly inferred from cluster luminosity functions. A 
list of nearby starburst systems is given in de Grijs et al.\ 2003; 
\nocite{2003MNRAS.343.1285D}
many cluster systems appear to have power law slopes of $\alpha \sim -2$.

Direct detection of cluster masses 
 via their stellar velocity dispersions and cluster sizes can be done for the closest systems, thereby giving direct determinations of cluster mass functions. Even in smaller
SSC systems, dynamical estimates of cluster masses are an important check
on mass functions obtained from cluster luminosity functions.
Masses at the high ends are consistent with globular cluster masses. 
In M82, McCrady \& Graham (2007) \nocite{2007ApJ...663..844M}
 find a power law mass distribution with slope
of $\alpha =-1.91$ for fifteen SSCs, nearly identical to the Antennae. 

An intriguing indication of ICMF evolution is the cluster system in 
NGC~5253. There are several hundred clusters in this galaxy
\citep{1989ApJ...338..789C}, many of intermediate age, $\sim$1 Gyr.
There is also a younger population of visible and embedded SSCs with ages
of $\sim$2--50 Myr 
\citep{1997AJ....114.1834C, 2001ApJ...554L..29G, 2002AJ....124..166A}. 
The cluster mass function
for the intermediate age clusters appears to turn over at a mass of 
$5\times 10^4$~M$_\odot$, while the younger SSCs have a power law
mass function \citep{2005A&A...433..447C}. 
Parmentier et al.\ (2008) \nocite{2008ApJ...678..347P} 
suggest that the evolution
of the ICMF is due to star formation efficiency (see Section~\ref{sec:7}).

Future work on stellar cluster mass functions in the coming decade
will be done from space with HST and JWST and
from the ground with IR AO observations. High resolution and sensitivity are
important for these studies, since fields where SSCs are found are usually
very crowded. Infrared observations are critical for getting the
mass functions of the youngest clusters, the best reflection of the ICMF,
since they are likely to be embedded in regions of high extinction. 

\section{What environments lead to extreme star formation?}
\label{sec:6}

\begin{figure}[b]
\begin{center}
\includegraphics[scale=.5]{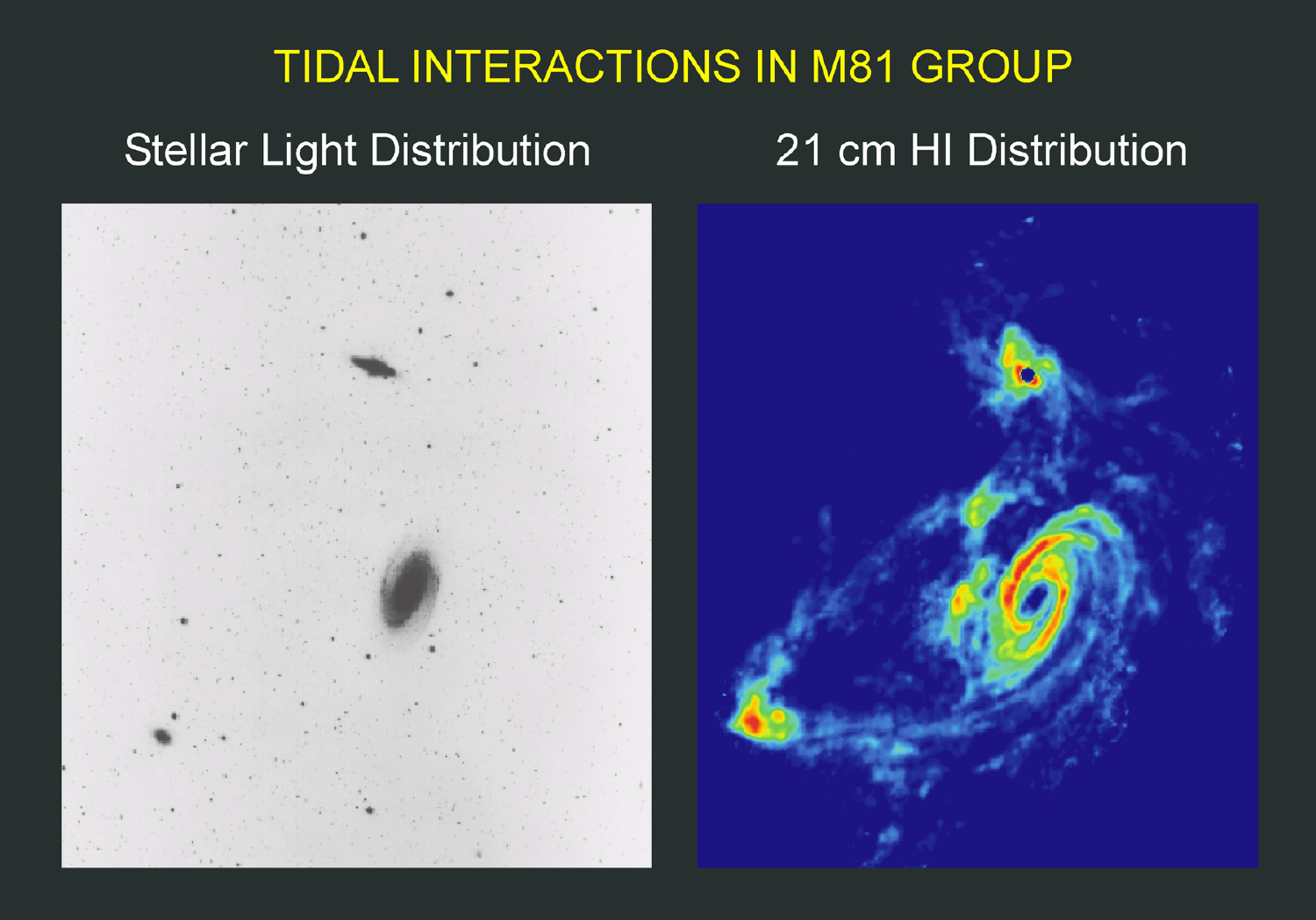}
\caption{The M81-M82-NGC~3077 group. (Left) Palomar Sky Survey Image. 
(Right) VLA image of the 21cm line of HI. Yun et al.\ (1994). Credit: NRAO/AUI/NSF.} \label{fig:2} 
\end{center}      
\end{figure}

\nocite{1994Natur.372..530Y}
In Figure~\ref{fig:2} 
is shown the nearby M81-M82-NGC~3077 interacting group of galaxies
\citep{1994Natur.372..530Y}. On the left is the Palomar Sky survey image, showing the 
stellar luminosity; on the right, is the VLA mosaic
of the group in the 21 cm line of HI. The ties that bind this group are obvious
in the 21 cm line emission. M82 is one of the best-known starburst galaxies, 
 with $L_{IR}\sim 6\times 10^{10}$~L$_\odot$,
and an estimated  $10^4$--$10^5$ O stars ($\rm N_{Lyc}\sim 8\times 10^{54}~s^{-1}$). NGC~3077 also has a modest starburst, of 
$\rm L_{IR}\sim 3 \times 10^8$~L$_\odot$ 
\citep[$\rm N_{Lyc}\sim 2\times 10^{52}~s^{-1}$:][]{2003AJ....126.1607S}, 
and M81 has a mildly active nucleus.
Clearly the conditions for ``extreme" star formation are favorable in this group. 
The atomic hydrogen of this interacting group has its own history, which is different from that of 
 the stars within the galaxies; there is evidence that some of the starburst activity is caused
 by a delayed ``raining down" of orbiting gas onto the galaxies several 
Myr after their closest encounters \citep{2001AJ....122.1770M}.
 
 The most luminous
infrared galaxies in the universe, with $\rm L_{IR} > 10^{11}$~{L$_\odot$}, are merging and interacting
systems, and these tend to be systems dominated by star formation. Our knowledge of the
stellar distributions is greater than our knowledge of the gas:
potentially many groups of galaxies have the connected appearance of the M81 group
with tidal loops in atomic hydrogen, and perhaps even in molecular gas, since starburst
galaxies are especially rich in molecular gas \citep{1989ApJ...340L..53M}. 
Arrays with wide-field capability, such as the Allen
Telescope Array, the Green Bank Telescope, and the SKA, are well-suited to the mapping of large
HI fields in nearby galaxy groups.

Star formation efficiency (SFE) is an important characteristic of the star 
formation process, since it is a measure of how efficiently molecular clouds are
turned into stars. SFE is closely tied to the ``infant mortality" of SSCs described in the next
section. There are numerous measures of the efficiency of star formation, broadly defined 
as the proportion of star formation per unit gas.
The Schmidt law, in which stars follow a power law correlation
with density \citep{1959ApJ...129..243S}, or the corresponding Kennicutt law, in terms of 
gas surface density \citep{1998ARAA..36..189K, 1998ApJ...498..541K} show
that star formation on global scales in galaxies is correlated with gas density.
The observable $\rm L_{IR}/M(H_2)$ is also used as an indication
of SFE.

While from a global perspective the total gas content, H{\sc I} + H$_2$, of galaxies
appears to be well correlated
 with star formation tracers \citep{2002ApJ...569..157W, 2007AJ....134.1827C}, 
observations of star-forming regions in the Galaxy indicate that stars
form from molecular gas clouds, rather than atomic. 
CO observations show a good correlation of $\rm L_{CO}$ with $\rm L_{IR}$ 
\citep{1986ApJ...304..443Y, 1986ApJ...305L..45S, 1991ARA&A..29..581Y, 
1996AJ....112.1903Y}. 

\begin{figure}[t]
\begin{center}
\includegraphics[scale=1.3]{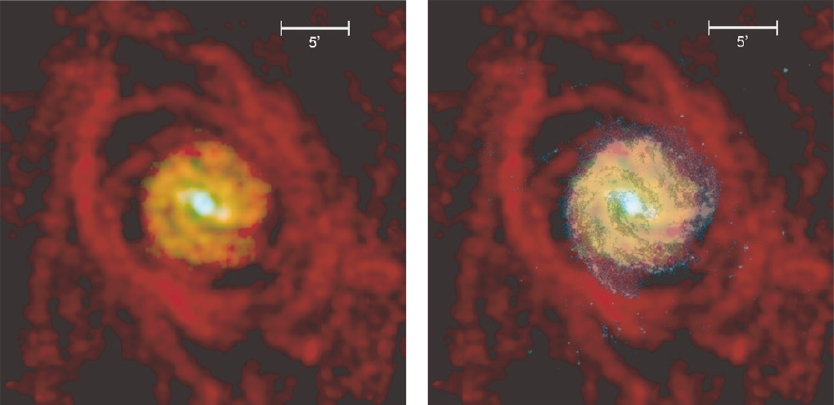}
\caption{(Left) The barred spiral galaxy M83. Red is a VLA image
of HI 21 cm line emission, Tilanus and Allen
(1993), and yellow is CO emission mapped with the NRAO 12 Meter
Telescope, Crosthwaite et al.\ (2002).  
(Right) Neutral gas with GALEX image overlay; GALEX image,  
Thilker et al.\ (2005), Gil de Paz et al.\ (2007).}\label{fig:3}
\end{center}       
\end{figure}

The atomic and molecular gas distribution in the spiral galaxy M83 is
shown in Figure~\ref{fig:3} 
 along with an overlay of the \textsc{GALEX}
ultraviolet image. This figure illustrates the general result
that while the atomic hydrogen disk can far exceed the visible disk of
a spiral galaxy, the optical portion of a spiral galaxy is primarily molecular gas.
Gas that forms stars is molecular. While a good correlation of
star formation tracers is seen with CO emission, 
an even tighter correlation is seen between star formation and the
dense ($\rm n>10^{5-6}~cm^{-3}$)
gas tracer HCN \citep{2004ApJ...606..271G}. This should not be surprising,
since denser gas is more likely to form stars.

What complicates the study of molecular gas in star-forming regions
is the necessity of using proxies, generally CO, to map out the distribution of $\rm H_2$.
A conversion factor between CO line intensity and $\rm H_2$ mass seems
to work well in the Galaxy, but will it do as well in ESF environments?

\nocite {1993A&A...274..707T} \nocite{2002AJ....123.1892C} 
\nocite{2005ApJ...619L..79T}  \nocite {2007ApJS..173..185G} 

Because of high energies of its first excited states, H$_2$ tends to be in
the ground state for temperatures less than 100K. Most Galactic giant
molecular clouds (GMCs)
have temperatures of 7--12K \citep{1985ApJ...289..373S,1987ApJS...63..821S},
although clouds in starbursts can be warmer. By contrast,
CO is relatively abundant, easily excited and thermalized, with a lowest
energy level equivalent temperature $E/k \sim 5.5$K.
For these reasons, CO lines are generally very optically thick. 
Yet CO is observed to be a good tracer of mass \citep{1987ApJ...319..730S}. 
This is because 
Galactic disk GMCs appear to be  turbulently supported against gravity, and
near virial equilibrium \citep{1983ApJ...270..105M}.
GMCs, which consist of optically thick clumps with turbulent motions larger than
systematic motions, have line profiles are Gaussian in spite of
high optical depths, and ``Large Velocity Gradient" (Sobolev approximation)
radiative transfer holds \citep{1993ApJ...402..195W}. 
The empirically-determined Galactic conversion factor, 
$\rm X_{CO} =\rm N_{H2}/I_{CO},$ is thus a dynamical
mass tracer  \citep{1986ApJ...309..326D,1987ApJ...319..730S}, 
in effect a Tully-Fisher relation for molecular clouds. 
Gamma ray observations indicate that a conversion factor of 
$\rm X_{CO} = 1.9 \times 10^{20}~\rm cm^{-2}\,(K~km\,s^{-1})^{-1}$ 
predicts $\rm H_2$ mass to within a factor of two within the Galaxy, 
with some radial variation \citep{1988A&A...207....1S, 2004A&A...422L..47S}.  
As an indicator of dynamical mass, $X_{CO}$ may actually be 
more robust than optically thin gas tracers in extreme
environments, since mass estimates based on optically thin
tracers depend upon temperature and relative abundance \citep{1988ApJ...325..389M,
2001ApJ...547..792D}. 

While the CO conversion factor seems to 
work well in the Galaxy, and as a dynamical mass tracer may be more robust
than tracers that are abundance-dependent, the association of CO and $\rm H_2$ 
has not been extensively tested in extreme environments. There
are clearly some situations  in which $X_{CO}$ fails to work well. 
The Galactic value of $X_{CO}$ does not give good masses
for the gas-rich centers of ultraluminous galaxies. In Arp 220, it overestimates
the mass by a factor of $\sim$5 due to a gas-rich nucleus, consisting of two
counterrotating disks \citep{1999ApJ...514...68S}, 
which are dominated by a warm, pervasive 
molecular gas  in which systematic motions dominate over turbulence 
\citep{1993ApJ...414L..13D, 1997ApJ...478..144S, 1998ApJ...507..615D}.  
$X_{CO}$ also appears to overpredict $\rm H_2$
masses in the  centers of local gas-rich spiral galaxies, including our own 
\citep{1998A&A...331..959D} by factors of 3--4.  CO 
appears to systematically misrepresent H$_2$ mass in spiral galaxies  
when the internal
cloud dynamics may be different from Galactic disk clouds, such as in the
nuclear regions  where
tidal shear visibly elongates clouds, causing systematic cloud motions to dominate 
\citep{2004AJ....127.2069M, 2008ApJ...675..281M}. It may also fail where cloud
structure may fundamentally differ from Galactic clouds, 
as in the LMC \citep{1986ApJ...303..186I}, 
where magnetic fields may be dynamically less important \citep{2007A&A...471..103B}.

For an understanding of
the links between star formation and molecular gas in ESF regions we
 require improved confidence in molecular gas masses in environments that are
 atypical of the Galaxy.  Systematic studies of molecular clouds in different tracers
 of molecular gas, 
 including dust, in ESF galaxies at high resolution in the millimeter and submillimeter 
with ALMA, CARMA, Plateau de Bure, and SMA
 will shed light on when we can confidently use CO as a tracer of molecular
 gas mass, and under 
what conditions it ceases to be a good tracer. 

Environmental factors other than gas mass are also important in the
fostering of star formation, but these are not as yet well understood.
\citet{1990ApJ...352..595T} concluded that the
efficiency of massive star formation is
higher in spiral arms than between the arms, consistent
with the ``strings of pearls along the spiral arms" description of nebulae by 
\citet{1957Obs....77..165B}. Clearly star formation is enhanced by spiral
arms, but how their large-scale influence trickles down in a 
turbulent GMC to a
parsec-scale core is not at all clear \citep{2007ApJ...661..972P}. 
Starburst rings also appear to facilitate star formation, particularly 
young SSCs \citep{1995AJ....110.1009B, 1996ApJS..107..215M, 2001AJ....121.3048M}.
SFE appears to steadily increase with the ferocity of the star formation.
While $\rm L_{IR}/M_{H_2}\sim 4$~L$_\odot$/M$_\odot$ for the Galaxy,
 it is $\sim$10--20~L$_\odot$/M$_\odot$ in regions of active star formation,
$\sim$20--100 in ULIRGs \citep{1986ApJ...305L..45S}. These studies
rely on $L_{IR}$, which may have contributions from older stellar populations.
Studies of star formation efficiency using tracers of recent star formation are
ongoing. 

\begin{figure}[t]
\begin{center}
\includegraphics[scale=.7]{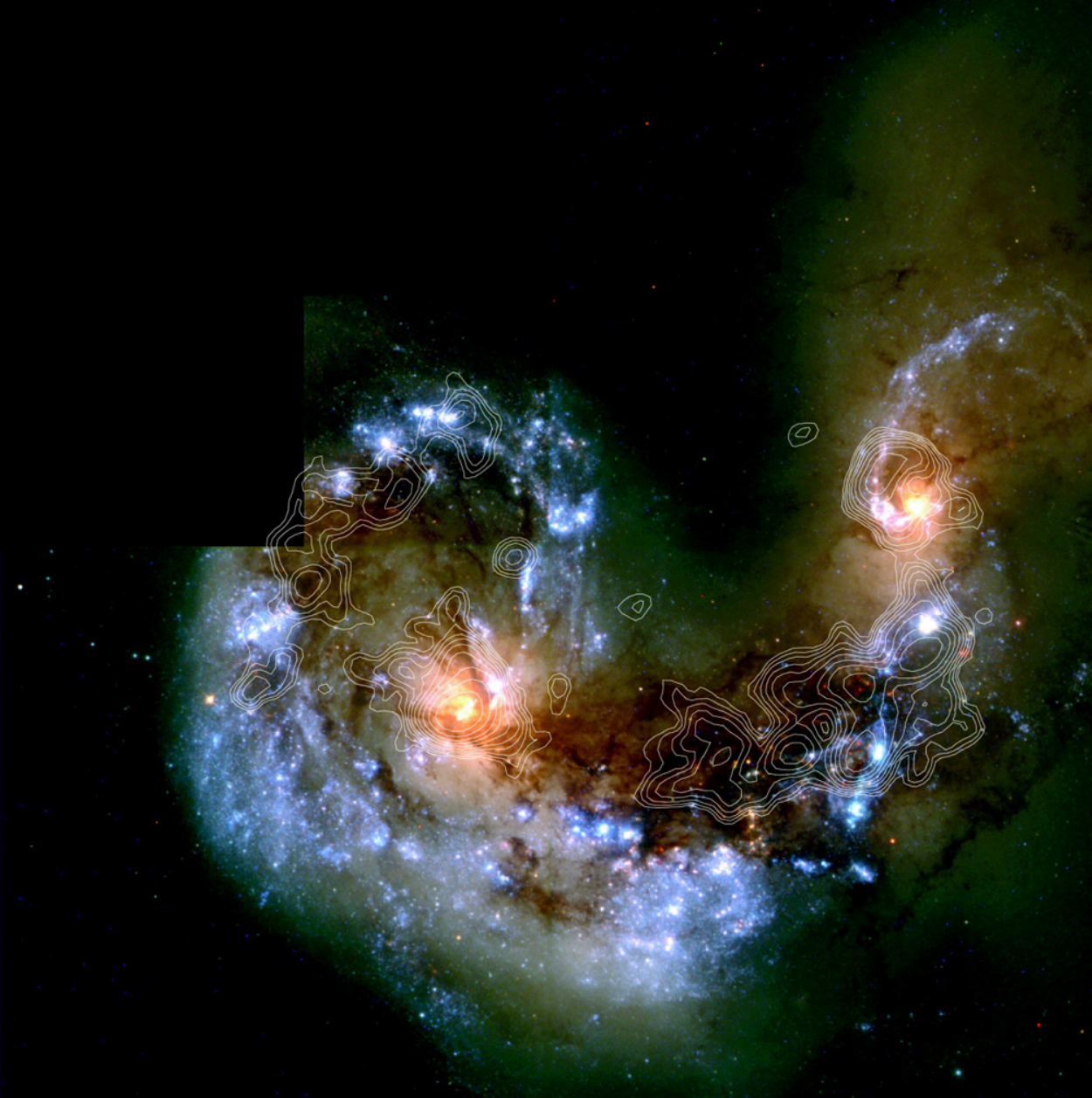}
\caption{CO in the Antennae. Contours are emission in the J=1--0 line of CO at 3mm,
imaged with the Owens Valley Millimeter Array. In color is the HST image.  
Wilson et al.\ (2000), Whitmore et al.\ (1999). Credit: C. Wilson.}\label{fig:4} 
\end{center}      
\end{figure}

Extreme star formation should require extreme amounts of molecular gas, and
luminous infrared galaxies have plenty of it \citep{1985ApJ...298L..31S, 
1986ApJ...311L..17Y, 1993ApJ...414L..13D, 1996ARA&A..34..749S}.
The Antennae interacting system, with its thousands of young SSCs,  contains 
exceptional amounts of gas. A CO image made with the Owens Valley Millimeter Array
by Wilson et al. (2000, 2003) \nocite{2000ApJ...542..120W, 2003ApJ...599.1049W} 
is shown on the HST image of \citep{1999AJ....118.1551W} in Figure~\ref{fig:4}. 
The greatest concentration of molecular gas is in the dusty, obscured region between
the two galaxies where the brightest infrared emission is found \citep{1998A&A...333L...1M}. There is an estimated $10^{9}$~M$_\odot$ of molecular gas in the
Antennae, supporting a current star formation
rate of $\rm N_{Lyc}\sim 10^{54}\, s^{-1}$  (Stanford et al.\ 1990, for D=13 Mpc, 
and $X_{CO} \sim 2\times 10^{20}~\rm cm^{-2}\,\,K~km\,s^{-1}$). 
\nocite{1990ApJ...349..492S} 
Based on the total mass of K-band selected clusters, 
the SFE is about 3--6\%\ \citep{2005A&A...443...41M},
 slightly higher than Galactic efficiencies on these scales
\citep{1984ApJ...285..141L} but not by much. $\rm H_2$ emission observed
using Spitzer in the dusty collision region suggests that ``pre-starburst" shocks may trigger the
star formation there \citep{2005A&A...433L..17H} as seen in other interacting
ULIRGS \citep{2006ApJ...648..323H}. This is a 
good illustration of the ``Super Giant Molecular Clouds" posited for interacting systems by 
\citet{1994ApJ...429..177H, 2002IAUS..207..545H}. Like the stellar cluster mass function,
which is power law with $\alpha \sim -2$, the mass function of 
giant molecular clouds in the Antennae is also power law, but with a slightly
different slope, $\alpha \sim -1.4$ \citep{2003ApJ...599.1049W}.

\nocite{1999AJ....118.1551W} \nocite{2000ApJ...542..120W} 

A counterexample to the ``lots of gas, lots of stars" theory is the case of NGC~5253. 
In this starburst dwarf galaxy, the star formation efficiency appears to be 
extremely high, SFE $\sim 75$\% on 100 pc scales, with $\rm H_2$ masses based
on both CO \citep{1997ApJ...474L..11T, 2002AJ....124..877M} and dust emission
(Turner et al.\ 2008, in prep.)  This SFE is nearly two orders of magnitude 
higher than generally seen on GMC spatial scales in the Galaxy.
NGC~5253 has several hundred young clusters, including several SSCs 
\citep{1996AJ....112.1886G,1997AJ....114.1834C}. 
Why is this galaxy so parsimonious in its usage of gas compared to
the Antennae? One difference between this starburst  and that of 
the Antennae is that NGC 5253 is a dwarf galaxy, with an estimated
mass of $\sim 10^{9}$~M$_\odot$, which was probably
originally a gas-poor dwarf spheroidal galaxy that has accreted
some gas \citep{1989ApJ...338..789C}. Unlike the Antennae, 
which are in the midst of a full-blown major
merger, NGC~5253 is relatively isolated, although part of the Cen A--M83
group \citep{2007AJ....133..504K}. The prominent dust lane entering the minor
axis  is the  probable cause of the starburst. Molecular gas is present in the dust lane, 
and this gas is observed to be falling into the
galaxy near the current starburst \citep{2002AJ....124..877M, 2008AJ....135..527K}. 
Radio recombination
line emission imaged at high resolution with the VLA of the central ``supernebula"
 shows a velocity gradient in the same direction as that of the
infalling streamer  \citep{2007ApJ...670..295R}. 
High star efficiency  is a necessary condition that the SSCs can evolve into globular clusters 
\citep{1997MNRAS.286..669G}. NGC~5253 may be the best case yet for
a galaxy in which the clusters might survive to become globular clusters.

The high resolution and sensitivity of ALMA, the new CARMA array, Plateau de
Bure and SMA will allow many more such starburst systems to be imaged in molecular
lines for study of the variation in SFE with environment. 
Star formation efficiency is a key parameter in the formation of long-lived 
clusters.

\section{Super Star Cluster Mortality}
\label{sec:7}

\begin{figure}[t]
 \begin{center}
\includegraphics[scale=.45]{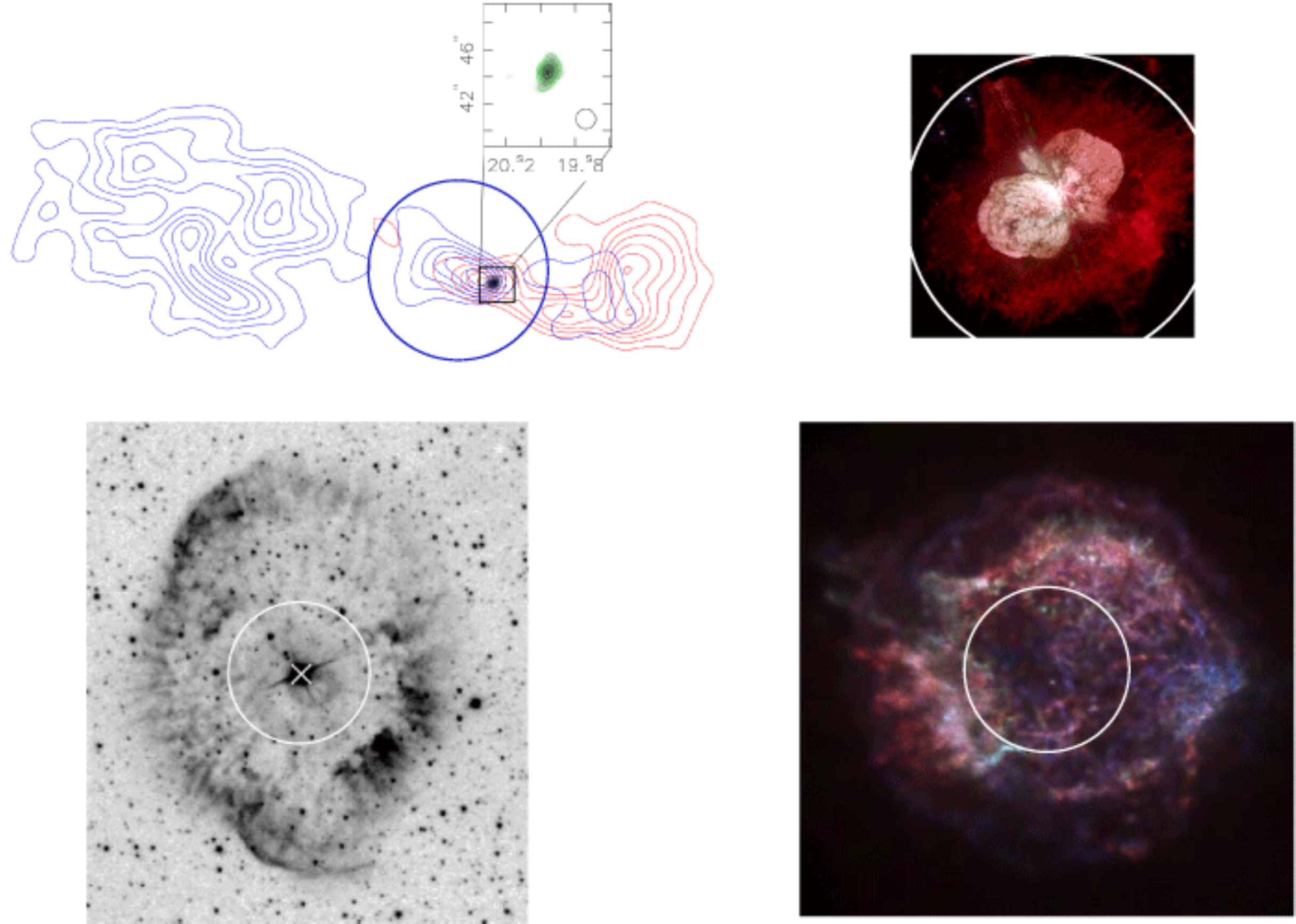}
\caption{The many ways that O stars can be destructive. Circles represent a region
1 pc across, the size of the core of an SSC. (top left)  Owens Valley
Millimeter Array image of the CO outflow source
around the massive protostar G192.16-3.82, Shepherd and Kurtz (1999). 
(top right) The Homunculus Nebula in Eta Carina imaged by HST,  Morse et al.\ (1998). 
(lower left) Wolf-Rayet bubble RCW58 in H$\alpha$, Gruendl et al.\ (2000). 
(lower right) Chandra image of the 1000-yr-old SNR Cas A, Hughes et al. (2000).}
\label{fig:5} 
 \end{center}      
\end{figure}

Birth and death go hand in hand in young SSCs, since O stars barely stop
accreting before they die \citep{2007ARA&A..45..481Z}. 
During their short lifetimes, O stars find many different
ways to lose mass. Windy and explosive by nature, O stars emit copious numbers of destructive
ultraviolet photons and are responsible for much of the mass loss and mechanical
feedback within a star cluster. 
Outflows, wind bubbles, LBV mass ejections, and SNR from a single O star can
influence a region the size of a young SSC; imagine what an 
SSC consisting of thousands of O stars can do to a parsec-scale volume!
Figure~\ref{fig:5}
 illustrates a few of the many ways that O stars can
destructively interact with their environments:  
CO outflows in adolescence (1~Myr); LBV outflows similar to
that responsible for 
 the Homunculus Nebula in early adulthood (2--3~Myr); Wolf-Rayet wind bubbles
at retirement (3--4~Myr); death by supernova (5--10~Myr). If the young cluster
manages to survive through the paroxysms of its riotous O star siblings, then there is 
most likely a molecular cloud nearby to unbind it. 

There is strong evidence that most  SSCs in the local universe will not survive
to old age. The odds of survival even  in the relatively benign environments of
the Galactic disk and halo are slim. Only an estimated $\sim$7\% 
 of embedded young clusters in the solar neighborhood will
live to the age of the Pleiades \citep{2003ARA&A..41...57L}. Dynamical models
of the Galactic globular cluster system suggest that as much as 75\%\ of 
the Galactic stellar halo may have originally been in the form of globular clusters
\citep{1997ApJ...474..223G, 2008MNRAS.386L..67S}, 
which now account for only 1\% of visible halo stars \citep{1998ASPC..136...33H}. 

The first hurdle that a young SSC must overcome is star formation efficiency,
defined here as 
$\eta = M_{stars}/(M_{stars}+M_{gas})$.
 The gas mass contribution includes contributions from ionized gas 
and atomic and molecular gas from the natal clouds.
Nominally $\eta>50$\%\ is required to leave a bound cluster 
\citep{1980ApJ...235..986H,1983ApJ...267L..97M,1984ApJ...285..141L}.
Lower values of $\eta \sim 30$\% can be accommodated if the cluster loses stars,
but retains a smaller bound core, leading to smaller clusters. 
The cluster may also survive if gas is lost quasistatically, so that
the cluster adjusts to the new equilibrium \citep{2002MNRAS.336.1188K}; 
it may also survive, although
with a significantly reduced stellar mass, if it suffers rapid mass loss (``infant
weight-loss") early on \citep{2006MNRAS.369L...9B}. Models suggest that
clusters with masses less than $10^5$~M$_\odot$ lose their residual gas quickly,
and that 95\% of all clusters are so destroyed within a few tens of 
Myr, and that rapid gas expulsion may give a natural
explanation for the lognormal PDMF for globular clusters  \citep{2008MNRAS.384.1231B}.

Based on the SFE/$\eta$ as observed in the Galaxy, the picture looks bleak for young SSCs.
On the sizescales of giant molecular clouds, $\eta$ is at most a few percent
\citep{2003ARA&A..41...57L}, a number
that appears to be determined by the turbulent dynamics of clouds 
\citep{2002ApJ...576..870P, 2006ApJ...653..361K, 2007ApJ...661..972P}.
On smaller scales in the Galaxy, $\eta \sim 10$\%--30\% 
(NGC~3603; N{\"u}rnberger et al.\ 2002), \nocite{2002A&A...394..253N} 
which is still rather small
to preserve a significant bound cluster on globular cluster scales. 

\nocite{1999ApJ...523..690S} 
\nocite{2000AJ....120.2670G} 
\nocite{1998AJ....116.2443M} 
 \nocite{2000ApJ...528L.109H} 

There is both fossil and structural observational
evidence that clusters dissolve. The vast
majority of bright SSCs with masses over $10^5$~M$_\odot$ in 
known SSC systems are less than 10 Myr in age 
 \citep{2005A&A...443...41M,2005A&A...443...79B}. 
From STIS spectroscopy,
Tremonti et al.\ (2001) and Chandar et al.\ (2005) 
\nocite{2001ApJ...555..322T, 2005ApJ...628..210C} 
find that the field stars in the nearby starburst galaxy NGC 5253 can be modeled
by dispersed cluster stars, with cluster dissolution timescales of
7--10 Myr.  In the Antennae system, 
\citet{2005ApJ...631L.133F} find that
the number of clusters falls sharply with age, with $\sim$50\%\ of the
stars in clusters
having dispersed after 10 Myrs.  They estimate that at least 20\%\, and
perhaps all, of the disk stars in the Antennae have formed within clusters. 
On the other hand, there is fossil evidence, in the form of
globular clusters,  that large clusters can and do survive for
many Gyr.

Star formation efficiency and cluster disruption may imprint upon cluster mass
functions. Parmentier et al.\ (2008) \nocite{2008ApJ...678..347P}
suggest that at $\eta \sim 20$\%, a power-law
core mass function turns into a bell-shaped cluster mass function, while
at higher efficiencies the power law is preserved. 
Gieles \& Bastian (2008) \nocite{2008A&A...482..165G}
suggest that the maximum cluster mass and age is a diagnostic of cluster
disruption, and they see evidence in cluster mass function, that 
formation/disruption does vary among galaxies.

The question remains of how globular clusters have managed to live to
such a ripe old age. Can we identify SSCs forming today that might indeed live to
become 10 billion years old? What initial conditions favor SSC longevity?
Going to deeper limits in the cluster
luminosity function could illuminate the connection between today's
SSCs and the older globular cluster population \citep{2004ApJ...611..220C}.
This is an extremely active area of research, but there are currently
few examples of high resolution studies of the efficiency of star formation
on GMC sizescales in starbursts. ALMA will have the sensitivity and
resolution to enable these studies in many nearby galaxies.

\section{Radiative Feedback: Effects on Molecular Clouds and Chemistry}
\label{sec:8}

The starburst galaxy, M82,  has one of the earliest known
and best studied examples  of a galactic wind. Both mechanical
luminosity in the form of stellar winds and supernovae and radiative 
luminosity from starbursts are feedback
mechanisms that can potentially disrupt star formation and end
the starburst phase. Yet  there are galaxies, such as the Antennae,
observed to have thousands of SSCs spread over regions of hundreds
of pc extent,  with cluster ages spanning many
tens of Myr during tidal interactions lasting $\sim$100~Myr.  
Evidently, episodes of intense star
formation can take place over periods of time far longer
than the lifetimes of individual massive stars in spite of feedback.
The subject of galactic winds and feedback is a large and active
one, and has been recently reviewed by \citet{2005ARA&A..43..769V}. 
The effects of feedback on denser molecular gas in
ESF regions is not yet well characterized, and it is the 
molecular gas  from which the future generations of stars will form. 

Starburst feedback  
and star formation occur on different spatial scales.  Giant 
molecular clouds consist of clumps that are governed by turbulence; only a 
small fraction of these clumps contain cores, which are the
regions that collapse to form stars or star clusters
\citep{2007ARA&A..45..565M}. Current computational 
models of turbulent clouds can explain the canonical star formation 
efficiencies of a few percent as that fraction of turbulent clumps that become 
dense enough for gravity to dominate 
 \citep{2002ApJ...576..870P, 2006ApJ...653..361K, 2007ApJ...661..972P}.
Once a core begins to collapse, free fall is rapid and there is little
time for feedback to operate. It is more likely that feedback operates on longer-lived and 
lower density molecular cloud envelopes dominated by turbulence, 
but how this large scale effect communicates down to the
small and rapidly collapsing star forming cores is unclear
 \citep{2007ApJ...668.1064E,2007ApJ...654..304K,2007ApJ...661..972P}.
 If rich star clusters form stars for several dynamical times, the energetic input
feedback could become important \citep{2006ApJ...641L.121T}.

One surprising characteristic of the interstellar medium in regions of
ESF is the ubiquity of dense ($\rm n_e \sim 10^{4-5}~\rm cm^{-3}$), 
``compact" H{\sc II} regions, a stage of star formation that should be
fleeting and relatively rare.  First detected 
spectroscopically in dwarf galaxies, such as NGC 5253 and He~2--10
\citep{1998AJ....116.1212T, 1999ApJ...527..154K, 2000AJ....120..244B},
these nebulae are detected by their high free-free optical depths
at cm wavelengths. These are
the ESF analogs of dense Galactic ``compact" H{\sc II} regions, only much
larger in size because of the high Lyman continuum rates from these large clusters.  
If the expansion of H{\sc II} regions is governed
by classical wind bubble theory, then the dynamical ages implied
by the sizes of these H{\sc II} regions are extremely short, tens of thousands
of years. M82-A1 is a young SSC with both a visible H{\sc II} region and a cluster, in
which  the dynamical age of the H{\sc II} region is too small for the cluster age
\citep{2006MNRAS.370..513S}.  
The H{\sc II} region around the Galactic cluster NGC 3603 is also too small for 
the cluster age  \citep{1995AJ....110.2235D}.  
These may be scaled-up versions of the classic Galactic ultracompact H{\sc II}  problem, 
in which there are ``too many" compact H{\sc II} regions given their  inferred
dynamical ages \citep{1984ApJ...283..632D, 1989ApJS...69..831W}. 
A possible explanation for the long lifetimes of these
H{\sc II} regions  is confinement by the high pressure environment of nearby dense 
($n_{H_2}>10^5~\rm cm^{-3}$) and warm molecular clouds 
 \citep[e.g.,][]{1995RMxAA..31...39D, 2005ApJ...619..755D,
2006ApJS..167..177D}. Radiative cooling may provide 
an important energy sink for the output of young super star clusters 
\citep{2007ApJ...669..952S}. However, the standard wind-blown bubble theory that
is generally applied to the development of Galactic H{\sc II} regions 
\citep{1975ApJ...200L.107C, 1985Natur.317...44C}
may not always apply to the H{\sc II} regions surrounding SSCs, 
which are massive enough to exert a
non-negligible gravitational pull on their H{\sc II} regions
\citep{2002MNRAS.336.1188K}. The ``supernebula" in 
NGC~5253 appears to be gravity-bound, if not in equilibrium
\citep{2003Natur.423..621T}. Gravity could either stall the
expansion of an SSC H{\sc II} region's expansion or create a cluster wind akin
to a stellar wind, depending on conditions. It is clear 
that the high interstellar pressures in starburst regions are critical to their development,
and to the evolution of the nearby molecular clouds.

Molecular gas has a tremendous ability to absorb energy and radiate it away.  
This could explain the ability of galaxies to sustain starburst events of extended
duration such as the one that has produced the thousands of SSCs in the Antennae. 
Irradiation causes heating, ionization, dissociation, and
 pronounced chemical changes in molecular clouds. It also provides
 us with a rich spectrum of potential diagnostics of radiative feedback.

\begin{figure}[t]
\includegraphics[scale=.7]{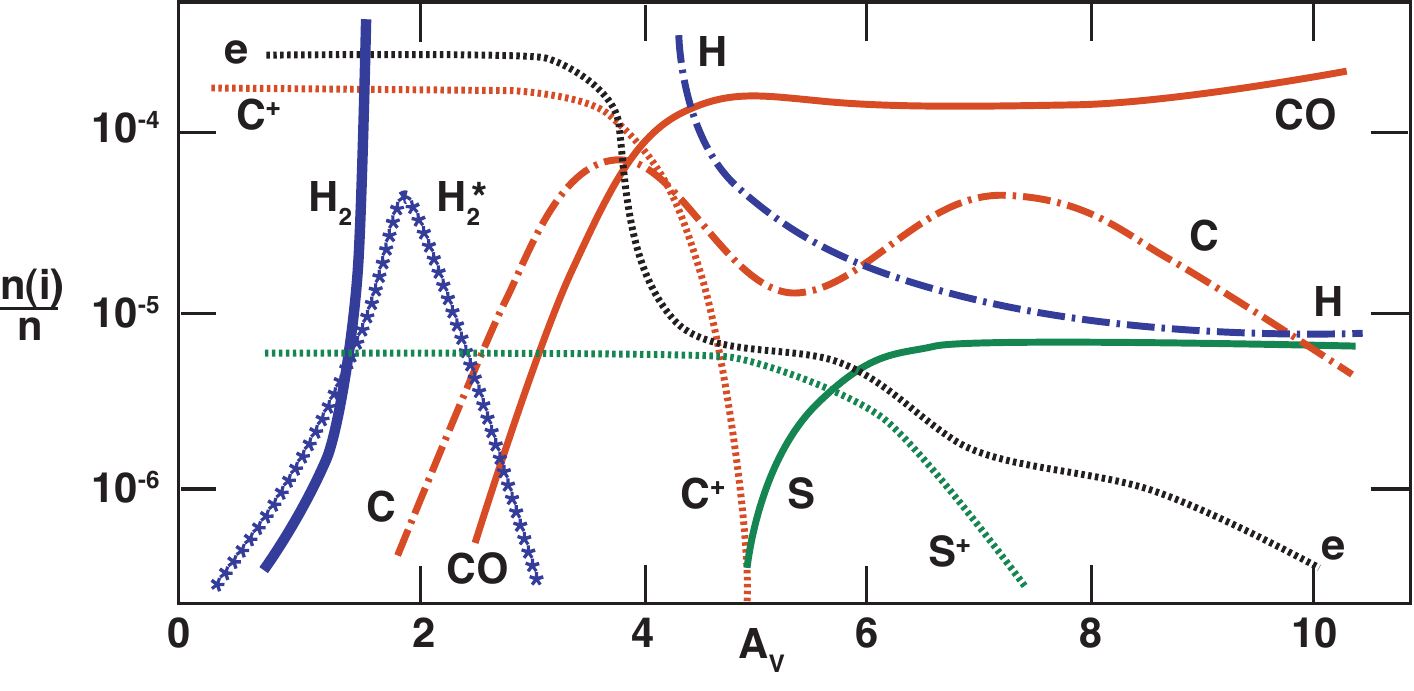}
\caption{Schematic of the ionization structure of the Orion photodissociation region (PDR), with relative elemental abundance plotted versus visual extinction. Adapted from Tielens \& Hollenbach (1985) and Tielens (2005).}
\label{fig:6}       
\end{figure}

\nocite{1985ApJ...291..722T} 
\nocite{2005pcim.book.....T}
In Figure~\ref{fig:6} 
 is shown a schematic of the ionization structure of a 
photodissociation region (PDR), adapted from \citet{1985ApJ...291..722T}.
(For a full description of PDRs, 
see  \citealt{2005pcim.book.....T} or \citealt{1999RvMP...71..173H} and 
also \citealt{1985ApJ...291..747T, 1996ApJ...458..222B, 1996ApJ...468..269D,
1999ApJ...527..795K, 2006ApJ...644..283K}.) 
Molecular clouds tend to
form $\rm H_2$ at $\rm A_V < \sim 1$. The translucent edge to the
molecular cloud can be quite warm, a few hundred K, warm enough
for excited $\rm H_2$ to be visible. 
Between $A_V=1$ and $A_V\sim 5$, while the clouds are molecular,
they also have significant abundances of heavy ions such as 
$\rm C^+$ and $\rm S^+$.
The presence of ions can drive a rich gas-phase chemistry through ion-molecule
reactions. The high temperatures can also liberate molecules that
have formed on the surfaces of grains in the form of ices in
the coldest clouds.  Warming the grains in either shocks or radiatively in PDRs
brings these molecules into the gas phase. PAH emission is also
bright from these regions, and dominates Spitzer images in the
8$\mu$m IRAC band where it shows a close association with star-forming
regions \citep{2004ApJ...613..986P, 2008ApJ...679..310G}. 
These chemical diagnostics have
been used to great effect in modeling the effects of protostars on their
surrounding protostellar disks, and in determining the shapes and
orientations of disks \citep{1998ARAA..36..317V}. Clouds near sources
of high X-ray radiation are subject to a similar phenomenon, but with
chemistry that is driven by a hard radiation field; these regions are called "XDRs" 
\citep{1996ApJ...466..561M, 1996A&A...306L..21L, 2005A&A...436..397M,
2006ApJ...650L.103M}.

 It might seem that processes that occur on scales of $\rm A_v\sim$1--5
would be difficult to detect in other galaxies, on the spatial 
scales of GMCs, but this is not the case.
Molecular clouds are porous, and there are many surfaces
within clouds of relatively low $\rm A_v$ individually; an estimated
90\% of molecular gas is in PDRs \citep{1999RvMP...71..173H}.  Some of the first
indications of the importance of PDR chemistry were the detections of
warm CO and the tracers of warmed, potentially shocked,
gas such as the C{\sc II} 158$\mu$m line  in starburst galaxies 
\citep{1991A&A...245..457M, 1991ApJ...373..423S}. 
Temperatures as high as 400--900K
have been inferred from lines of interstellar ammonia \citep{2003A&A...403..561M}.  

Lines of numerous heavy molecules have been detected from other galaxies,
and the brightest sources are the star-forming galaxies. Molecules such
as formaldehyde (H$_2$CO), methanol (CH$_3$OH), and
cyanoacetylene (HC$_3$N) have been detected in nearby starburst
galaxies \citep{2006A&A...450L..13M, 2008ApJ...673..832M, MeierTurner08}. Many of
these models can provide discriminants between PDR and XDR-heated
gas \citep{2006ApJ...650L.103M, 2007A&A...464..193A}. 
A spectral line survey of NGC~253 in the 2mm atmospheric
window reveals 111 identifiable spectral features from 25 different molecular
species; the spectrum suggests that the molecular abundances in NGC~253 
are similar to those of the Galactic Center, with a chemistry dominated by
low-velocity shocks \citep{2006ApJS..164..450M, 2008Ap&SS.313..303M}. 

\begin{figure}[t]
\begin{center}
\includegraphics[scale=1.1]{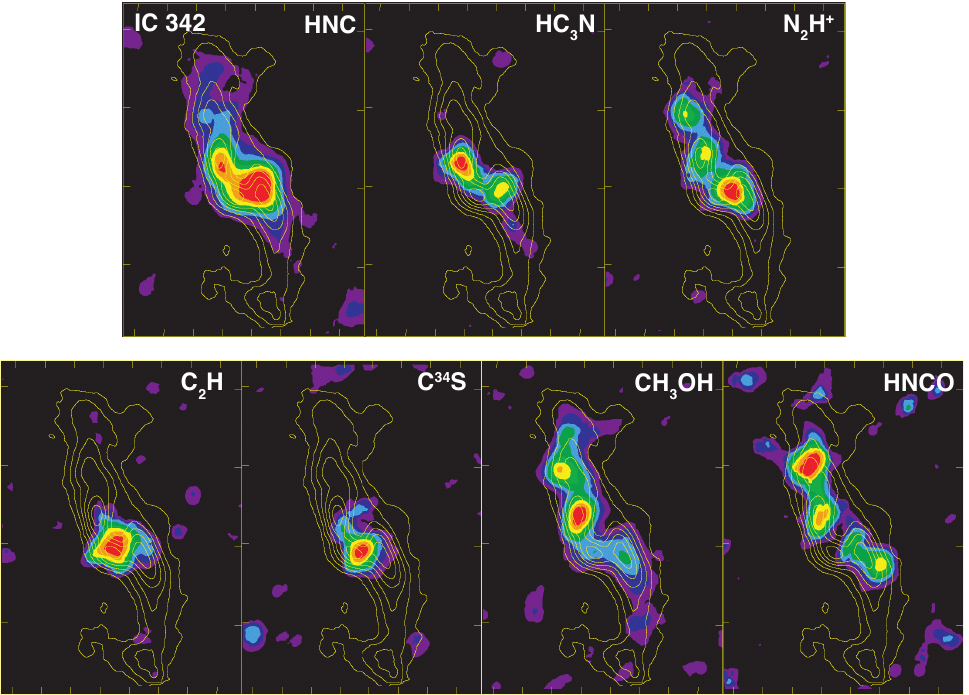}
\caption{Spatially-resolved (5$^{\prime\prime}$=75 pc) chemistry of the central 
300 pc of the Scd spiral
galaxy IC~342. Meier \& Turner (2005). Panels at top show molecules that are
overall molecular gas tracers. Bottom, left: C$_2$H and C$^{34}$S trace clouds in
 high radiation fields (PDRs); right:
 CH$_3$OH (methanol) and HNCO trace grain-chemistry
 along the arms of the minispiral.}\label{fig:7} 
\end{center}      
\end{figure}

Imaging adds another dimension to the molecular line
spectra.  In Figure~\ref{fig:7} 
are shown 
interferometric images in several molecules
 of the nuclear ``minispiral" of the nearby Scd galaxy, 
IC~342. The lines shown are all at $\lambda =$3mm,
 and have comparable excitation energies
and similar  critical densities. The images show a clear variation in cloud chemistry
and molecular abundances 
with galactic location. A principal component analysis \citep{2005ApJ...618..259M} 
shows that the molecules $\rm N_2H^+$, HNC, and HCN have similar
spatial distributions to CO and its isotopologues, and are good tracers of the
overall molecular cloud distribution. The molecule $\rm CH_3OH$ (methanol)
is well known from Galactic studies to be a ``grain-chemistry" molecule,
which forms on grain surfaces and is liberated by shocks or warm 
cloud conditions; here, methanol and HNCO follow the arms of the 
minispiral. Methanol and HNCO appear to be tracing the gentle shocks
of the gas passing through the spiral arms.  The final group of molecules,
$\rm C^{34}S$ and $\rm C_2H$ are found in the immediate vicinity (50 pc)
of the nuclear cluster; these molecules presumably reflect the intense
radiation fields of the nuclear star cluster.

The next decade will see a blossoming of molecular line studies of ESF. The
current state of molecular line work in galaxies has been recently reviewed
by Omont (2007), \nocite{2007RPPh...70.1099O} 
and also is represented by contributions in the volume
by Bachiller \& Cernicharo (2008). \nocite{2008ASS...B} 
Spectroscopy of ESF gas in galaxies with the APEX, ASTE, CARMA,
IRAM 30 m, Plateau de Bure, CARMA, SMA, Spitzer, and the VLA 
telescopes will continue to probe
the star-forming ISM in nearby galaxies through the next decade. Herschel
and SOFIA
will soon provide spectroscopy of the important 158$\mu$m line of C\textsc{II}
from PDRs in ESF galaxies. 
ALMA will add tremendous sensitivity, milliarcsecond resolution, 
km/s velocity resolution, access to
the southern hemisphere, and submillimeter capability, allowing us to
study extreme star formation and extreme star-forming gas
 and its effects on galaxies
in exquisite detail.

\begin{acknowledgement}
I am grateful to Nate McCrady, David S. Meier, and Andrea Stolte
 for their helpful discussions and comments, and to Xander Tielens for
his good humor and patience. 
\end{acknowledgement}

\end{document}